\documentclass[10pt, article, lettersize]{IEEEtran}
\usepackage{epsf,cite,amsmath,amscd,amssymb,graphicx,latexsym,multicol}
\newtheorem{proposition}{Proposition}

\begin{document}

\centerfigcaptionstrue
\bibliographystyle{ieeetr}

\title{\huge{A Game-Theoretic Approach to Energy-Efficient Modulation\\in CDMA Networks with Delay Constraints}\thanks{F. Meshkati was with
the Department of Electrical Engineering, Princeton University. He
is now with QUALCOMM Inc. This research was supported by the
National Science Foundation under Grant ANI-03-38807.}}
\author{Farhad Meshkati$^{*}$, Andrea J. Goldsmith$^{**}$, H. Vincent Poor$^{***}$, and Stuart C.
Schwartz$^{***}$\vspace{0.2cm}\\
$^{*}$QUALCOMM Inc.\\ San Diego, CA 92121
USA\\ meshkati@qualcomm.com\vspace{0.2cm}\\
$^{**}$Dept. of Electrical Engineering, Stanford University\\
Stanford, CA 94305 USA\\ andrea@ee.stanford.edu\vspace{0.2cm}\\
$^{***}$Dept. of Electrical Engineering, Princeton
University\\ Princeton, NJ 08544 USA\\
\{poor,stuart\}@princeton.edu\vspace{-0.5cm}}

\maketitle \thispagestyle{empty} \pagestyle{empty}

\begin{abstract}
A game-theoretic framework is used to study the effect of
constellation size on the energy efficiency of wireless networks for
M-QAM modulation. A non-cooperative game is proposed in which each
user seeks to choose its transmit power (and possibly transmit
symbol rate) as well as the constellation size in order to maximize
its own utility while satisfying its delay quality-of-service (QoS)
constraint. The utility function used here measures the number of
reliable bits transmitted per joule of energy consumed, and is
particularly suitable for energy-constrained networks. The
best-response strategies and Nash equilibrium solution for the
proposed game are derived. It is shown that in order to maximize its
utility (in bits per joule), a user must choose the lowest
constellation size that can accommodate the user's delay constraint.
Using this framework, the tradeoffs among energy efficiency, delay,
throughput and constellation size are also studied and quantified.
The effect of trellis-coded modulation on energy efficiency is also
discussed.
\end{abstract}


\section{Introduction}

Wireless networks are expected to support a variety of
applications with diverse quality-of-service (QoS) requirements.
Because of the scarcity of network resources (i.e., energy and
bandwidth), radio resource management is crucial to the
performance of wireless networks. Adaptive modulation has been
shown to be an effective method for improving the spectral
efficiency in wireless networks (see for example
\cite{WebbSteele,GoldsmithChua97,GoldsmithChua98,Yoo05}). However,
the focus of many of the studies to date has been on maximizing
the throughput of the network, and the impact of the modulation
order on energy efficiency has not been studied to the same
extent. In \cite{Cui05}, the authors have used a
convex-optimization approach to study modulation optimization for
an energy-constrained time-division-multiple-access (TDMA)
network.

Game-theoretic approaches to power control have recently attracted
considerable attention (see, for example, \cite{MeshkatiTCOM} and
the references therein). In this work, we study the effects of
modulation on energy efficiency of code-division-multiple-access
(CDMA) networks using a \emph{competitive} multiuser setting.
Focusing on M-QAM modulation, we propose a non-cooperative game in
which each user chooses its strategy, which includes the choice of
the transmit power, transmit symbol rate and constellation size,
in order to maximize its own utility while satisfying its QoS
constraints. The utility function used here measures the number of
reliable bits transmitted per joule of energy consumed, and is
particularly suitable for energy-constrained networks. Using our
non-cooperative game-theoretic framework, we quantify the
tradeoffs among energy efficiency, delay, throughput and
modulation order. While game-theoretic approaches to resource
allocation with delay QoS constraints have previously been studied
in \cite{MeshkatiISIT} and \cite{MeshkatiDelayTCOM}, this is the
first work that takes into account the effect of modulation.

The remainder of this paper is organized as follows. The system
model and definition of the utility function are given in
Section~\ref{systemmodel}. We first discuss our proposed power
control game without any delay constraints in Section~\ref{pcg}
and derive the corresponding Nash equilibrium solution. The
delay-constrained power control game is presented in
Section~\ref{pcgd} and the corresponding best-response strategies
and Nash equilibrium solution are derived. The analysis is
extended to coded systems in Section~\ref{pcgtcm}. Numerical
results and conclusions are given in
Sections~\ref{numericalresults}~and~\ref{conclusion},
respectively.

\section{System Model} \label{systemmodel}

We consider a wireless network in which the users' terminals are
transmitting to a common concentration point (e.g., a base station
or an access point). We define the utility function of a user as
the ratio of its goodput to its transmit power, i.e.,
\begin{equation}\label{eq2}
   u_k = \frac{T_k}{p_k} \ .
\end{equation}
Goodput is the net number of information bits that are transmitted
without error per unit time and is expressed as
\begin{equation}\label{eq3}
   T_k = R_k f(\gamma_k)
\end{equation}
where $R_k$ is the transmission rate, $\gamma_k$ is the output
signal-to-interference-plus-noise ratio (SIR) for user $k$, and
$f(\gamma_k)$ is the ``efficiency function" which represents the
packet success rate (PSR). We require that $f(0)=0$ to ensure that
$u_k=0$ when $p_k=0$. In general, the efficiency function depends on
the modulation, coding and packet size. Based on (\ref{eq2}) and
(\ref{eq3}), the utility function for user $k$ can be written as
\begin{equation}\label{eq4}
   u_k = R_k \frac{f(\gamma_k)}{p_k}\ .
\end{equation}
This utility function, which has units of \emph{bits/joule},
measures the number of reliable bits that are transmitted per joule
of energy consumed, and is particularly suitable for
energy-constrained networks.


Focusing on M-QAM modulation, in this work we study
non-cooperative games in which the actions open to each user are
the choice of transmit power (and possibly transmit symbol rate)
as well as the choice of \emph{constellation size}. For the M-QAM
modulation, the number of bits transmitted by each symbol is given
by $$b=\log_2 M .\footnote{Since there is a one-to-one mapping
between $M$ and $b$, we sometimes refer to $b$ as the
constellation size.}$$ We focus on \emph{square} M-QAM modulation,
i.e., $b\in\{2,4,6,\cdots\}$, since there are exact expressions
for the symbol error probability of square M-QAM modulation (see
\cite{GoldsmithBook}). We can easily generalize our analysis to
include odd values of $b$ by using an approximate expression for
the symbol error probability. Let us for now focus on a specific
user and drop the subscript $k$. Assuming a packet size of $L$
bits, the packet success rate for square M-QAM modulation is given
by
\begin{equation}\label{eq15}
P_{\textrm{success}}(b,\gamma)= \left( 1- \alpha_b Q(\sqrt{\beta_b
\gamma})\right)^{\frac{2L}{b}}
\end{equation}
where\vspace{-0.2cm}
$$\alpha_b = 2\left(1-2^{-b/2}\right)$$ and\vspace{-0.2cm} $$\beta_b=
\frac{3}{2^b-1}.$$ Here, $\gamma$ represents the symbol SIR and
$Q(\cdot)$ is the complementary cumulative distribution function of
the standard Gaussian random variable. Note that at $\gamma=0$, we
have $P_{\textrm{success}}~=~2^{-L}~\neq~0$. Since we require the
efficiency function to be zero at zero transmit power, we define
\begin{equation}\label{eq16}
f_b(\gamma)= \left( 1- \alpha_b Q(\sqrt{\beta_b \gamma})
\right)^{\frac{2L}{b}} - 2^{-L}.
\end{equation}
Note that $2^{-L}\simeq 0$ when $L$ is large (e.g., $L=100$).

\section{Power Control Game with M-QAM Modulation}\label{pcg}

We consider a direct-sequence CDMA (DS-CDMA) network with $K$ users
and express the transmission rate of user $k$ as
\begin{equation}\label{eq14}
R_k= b_k R_{s,k}
\end{equation}
where $b_k$ is the number of information bits per symbol and
$R_{s,k}$ is the symbol rate. Let us for now assume that users have
no delay constraints. We propose a power control game in which each
user seeks to choose its constellation size and transmit power in
order to maximize its own utility, i.e.,
\begin{equation}\label{eq14b}
    \max_{b_k , p_k} R_k \frac{f(\gamma_k)}{p_k}\ \ \ \textrm{for} \
    \ k=1,\cdots,K ,
\end{equation}
where $b_k \in \{2,4,6,\cdots\}$ and $p_k \in [0, P_{max}]$ with
$P_{max}$ being the maximum allowed transmit power. Throughout this
work, we assume $P_{max}$ is large.

\begin{figure}[t]
\begin{center}
 \leavevmode \hbox{\epsfysize=6cm \epsfxsize=7cm
\epsffile{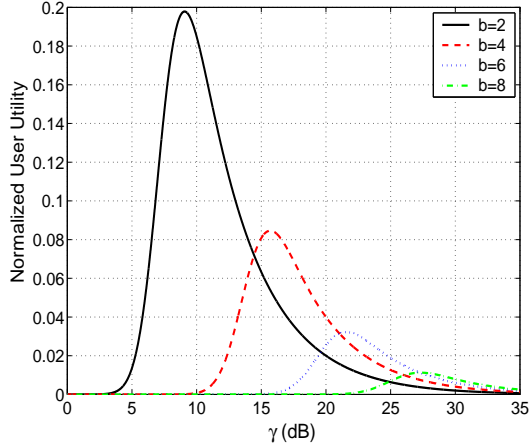}}\vspace{-0.2cm}
\end{center} \caption{Normalized user utility as a function of SIR for different constellation sizes.}\label{fig2}
\vspace{-0.2cm}
\end{figure}

For all linear receivers, the output SIR for user $k$ can be written
as\vspace{-0.1cm}
\begin{equation}\label{eq17}
    \gamma_k= (B/R_{s,k}) p_k\ \hat{h}_k
\end{equation}
where $B$ is the system bandwidth and $\hat{h}_k$ is the effective
channel gain which is independent of the transmit power and rate of
user $k$. Based on \eqref{eq14} and \eqref{eq17}, and by dropping
the subscript $k$ for convenience, the maximization in \eqref{eq14b}
can be written as
\begin{equation}\label{eq18}
    \max_{b , \gamma} B \hat{h}\ b
    \frac{f_{b}(\gamma)}{\gamma} .
\end{equation}
It is important to observe that, for a given $b$, specifying the
operating SIR completely specifies the utility function. Let us for
now fix the symbol rate $R_s$ and the constellation size. Taking the
derivative of \eqref{eq18} with respect to $\gamma$ and equating it
to zero, we conclude that the utility of a user is maximized when
its output SIR is equal to $\gamma_b^*$ which is the unique
(positive) solution of
\begin{equation}\label{eq19}
    f_b(\gamma) = \gamma f'_b(\gamma).
\end{equation}
The maximum utility is hence given by
\begin{equation}\label{eq20}
    u_b^*=B \hat{h}\ b
    \frac{f_{b}(\gamma_b^*)}{\gamma_b^*}
\end{equation}
We can compute $\gamma_b^*$ numerically for different values of $b$.
Table~\ref{table1} summarizes the results.
\begin{table}
\begin{center} \caption{}\label{table1}
\begin{tabular}{|c|c|c|c|c|c|c|c|c|}
  \hline
  $b$ & $\alpha_b$ & $\beta_b$ & $\gamma_b^*$(dB) & $f_b(\gamma_b^*)$ & $b/\gamma_b^*$(dB) & $b f_b(\gamma_b^*)/ \gamma_b^*$ \\
  \hline
  \hline
  2 & 1 & 1 &  9.1 & 0.801 & -6.1 & 0.1978\\
  4 & 1.5 & 0.2 &  15.7 & 0.785 & -9.7 & 0.0846\\
  6 & 1.75 & 0.0476 &  21.6 & 0.771 & -13.8 & 0.0322\\
  8 & 1.875 & 0.0118 &  27.3 & 0.757 & -18.3 & 0.0112\\
  10 &1.9375 & 0.0029 &  33.0 & 0.743 &-23.0 & 0.0037\\

  \hline
\end{tabular}
\end{center}
\end{table}
It is observed from Table~\ref{table1} that the user's utility is
maximized when $b=2$ (i.e., QPSK modulation). This is because, as
$b$ increases, the linear increase in the throughput is dominated
by the exponential increase in the required transmit power (which
results from the exponential increase in $\gamma_b^*$). As a
result, it is best for a user to use QPSK
modulation.\footnote{BPSK and QPSK are equivalent in terms of
energy efficiency, but QPSK has a higher spectral efficiency.}
Fig.~\ref{fig2} shows the normalized user utility (i.e.,
$\frac{u_b}{B \hat{h}}$) as a function of SIR for different
choices of $b$.

So far, we have shown that at Nash equilibrium (if it exists),
QPSK modulation must be used by each user. The existence of the
Nash equilibrium for the proposed game can be shown via the
quasiconcavity of each user's utility function in its own power.
Furthermore, it can be shown that the equilibrium is unique (see
\cite{MeshkatiModJournal} for more details).

\section{Delay-Constrained Power Control Game with M-QAM
Modulation}\label{pcgd}

We now extend the analysis of Section~\ref{pcg} to the case in
which the users have delay QoS requirements. Our goal in this part
is to study the effects of constellation size on energy efficiency
and delay. The delay QoS constraint considered here is in terms of
average delay.\footnote{Note that an average-delay constraint may
not be sufficient for applications with hard delay requirements
(see \cite{Holliday02}).}

Let us assume that the incoming traffic for user $k$ has a Poisson
distribution with parameter $\lambda_k$ which represents the average
packet arrival rate with each packet consisting of $L$ bits. The
user transmits the arriving packets at a rate $R_k= b_k R_{s,k}$
(bps) and with a transmit power equal to $p_k$ Watts. We assume an
automatic-repeat-request (ARQ) mechanism in which the user keeps
retransmitting a packet until the packet is received at the access
point without any errors. Also, the incoming packets are assumed to
be stored in a queue and transmitted in a first-in-first-out (FIFO)
fashion. The packet success probability (per transmission) as before
is represented by the efficiency function $f_b(\gamma)$.

Focusing on a specific user and dropping the subscript $k$, we can
represent the combination of the user's queue and wireless link as
an M/G/1 where the service rate, $\mu$, is given by
\begin{equation}\label{eq24}
    \mu= \frac{f_b(\gamma)}{\tau}=
    R_s \frac{b f_b(\gamma)}{L} .
\end{equation}
Here, $\tau$ represents the packet transmission time (i.e., $\tau
= \frac{L}{b R_s}$). Now, let $W$ be a random variable
representing the total packet delay (including transmission and
queuing delays) for the user. It can be shown that the average
packet delay is given by (see \cite{MeshkatiDelayTCOM} for
details)\vspace{-0.1cm}
\begin{equation}\label{eq25}
    \bar{W} = \tau \left(\frac{1-\frac{\lambda
    \tau}{2}}{f_b(\gamma)-\lambda \tau}\right) \ \ \
    \textrm{with} \ f_b(\gamma)>\lambda \tau .
\end{equation}
The delay QoS constraint for a user is specified by an upper bound
on the average packet delay, i.e., we require
\begin{equation}\label{eq26}
    \bar{W} \leq D .
\end{equation}
This delay constraint can equivalently be expressed as
$$\gamma\geq \hat{\gamma_b}$$
where\vspace{-0.2cm}
\begin{equation}\label{eq29}
    \hat{\gamma}_b=f_b^{-1}(\eta_b) \ .
\end{equation}
with\vspace{-0.2cm}
\begin{equation}\label{eq28}
    \eta_b = \frac{L \lambda}{b R_s}  + \frac{L}{b R_s D} -
    \frac{L^2 \lambda }{2 b^2 R_s^2 D} .
\end{equation}
This means that the delay constraint in \eqref{eq26}
translates into a lower bound on the output SIR.

We propose a game in which each user chooses its transmit power and
symbol rate as well as its constellation size in order to maximize
its own utility while satisfying its delay requirement. Fixing the
other users' transmit powers and rates, the best-response strategy
for the user of interest is given by the solution of the following
constrained maximization:\vspace{-0cm}
\begin{equation}\label{eq30}
    \max_{p, R_s, b} \ u \ \ \ \textrm{s.t.} \ \ \ \bar{W} \leq D \ ,
\end{equation}\vspace{-0.2cm}
or equivalently
\begin{equation}\label{eq31}
    \max_{\gamma, R_s, b} \ b \frac{f_b(\gamma)}{\gamma} \ \ \
     \textrm{s.t.} \ \  \gamma \geq \hat{\gamma}_b \ \ \textrm{and} \ \ 0\leq \eta_b<1 \
     .
\end{equation}\vspace{0cm}
%
Let us define $\Omega_b^*=\left(\frac{L}{D}\right)
\frac{1+D\lambda +\sqrt{1+ D^2 \lambda^2 +2(1-f_b^*)D
\lambda}}{2f_b^*}$,
where $f_b^*=f_b(\gamma_b^*)$. \vspace{0.1cm}
\begin{proposition} \label{prop2}
For given values of $\lambda$ and $D$, the best-response strategy
for a user (i.e., the solution of \eqref{eq30}) is any combination
of $p$ and $R_s$ such that
\begin{equation}
\min\left\{\Omega_{\tilde{b}}^*/\tilde{b}, B\right\} \leq R_s \leq B
\end{equation}\vspace{-0.2cm}
and\vspace{-0.2cm}
\begin{equation}\label{eq31c}
\gamma=\left\{
         \begin{array}{ll}
           \gamma_{\tilde{b}}^*, & \hbox{if  $\Omega_{\tilde{b}}^*/\tilde{b} \leq B$;} \\
           \hat{\gamma}_{\tilde{b}}, & \hbox{if  $\Omega_{\tilde{b}}^*/\tilde{b} > B$,}
         \end{array}
       \right.
\end{equation}
where $\tilde{b}$ is the lowest constellation size for which
$\lambda$ and $D$ are feasible, $\gamma_b^*$ is the solution of
\eqref{eq19}, and $\hat{\gamma}_{b}$ is given by \eqref{eq29}.
\end{proposition}
\begin{proof}
The proof is omitted due to space limitation
(see~\cite{MeshkatiModJournal}).
\end{proof}

Proposition~\ref{prop2} implies that, in terms of energy efficiency,
choosing the lowest-order modulation (i.e., QPSK) is the best
strategy unless the user's delay constraint is too tight. In other
words, the user would jump to a higher-order modulation only when it
is transmitting at the highest symbol rate (i.e., $R_s=B$) and still
cannot meet the delay requirement. Also, the proposition suggests
that if $\Omega_{\tilde{b}}^*/\tilde{b} < B$, the user has
infinitely many best-response strategies.

It can be shown that for a matched filter, the utility of user $k$
at Nash equilibrium is given by
\begin{equation}\label{eq33}
    u_k=\frac{B f(\gamma_k) h_k}{\sigma^2 \gamma_k}\left(1-
\frac{\sum_{j \neq k} \Phi_j}{1-\Phi_k}\right) \ ,
\end{equation}
where $\Phi_k=\left(1+\frac{B}{R_{s,k}\gamma_{k}}\right)^{-1}$ for
$k=1,\cdots,K$\footnote{$\Phi_k$ here is a generalized version of
the ``size" of user $k$ as defined in \cite{MeshkatiDelayTCOM}.}.
This implies that while our proposed game could potentially have
infinitely many Nash equilibria, the Nash equilibrium with the
smallest $R_{s,k}$'s achieves the largest utility. This means the
Nash equilibrium with
$R_{s,k}=\min\{\Omega_{\tilde{b}_k}^*/\tilde{b}_k, B\}$ for
$k=1,\cdots,K$ is the \emph{Pareto-dominant} Nash equilibrium.

\section{Power Control Games with Trellis-Coded M-QAM Modulation}
\label{pcgtcm}

In this section, we extend our analysis to trellis-coded
modulation (TCM). Let $G$ represent the effective \emph{coding
gain} achieved by  TCM as compared to the equivalent uncoded
system (see \cite{Ungerboeck}). In general, the coding gain is a
function of both the operating SIR and the modulation level.
Hence, the efficiency function for TCM is given by
\begin{equation}\label{eq34}
f_b^{(c)}(\gamma)\simeq \left( 1- \alpha_b Q\left(\sqrt{\beta_b
\gamma G_b(\gamma)}\right) \right)^{\frac{2L}{b}} - 2^{-L} ,
\end{equation}
where $b$ is the number of \emph{information} bits per symbol. One
can follow the same analysis for the coded system as the one
presented for the uncoded system by replacing $f_b(\gamma)$ with
$f_b^{(c)}(\gamma)$ given in \eqref{eq34}. Due to space
limitation, we omit the analysis (see \cite{MeshkatiModJournal}
for more details). We will show in Section~\ref{numericalresults}
that as expected, for the same spectral efficiency, the energy
efficiency is higher when TCM is used.

\section{Numerical Results}\label{numericalresults}

\begin{figure}
\begin{center}
 \leavevmode \hbox{\epsfysize=8.5cm \epsfxsize=9cm
\epsffile{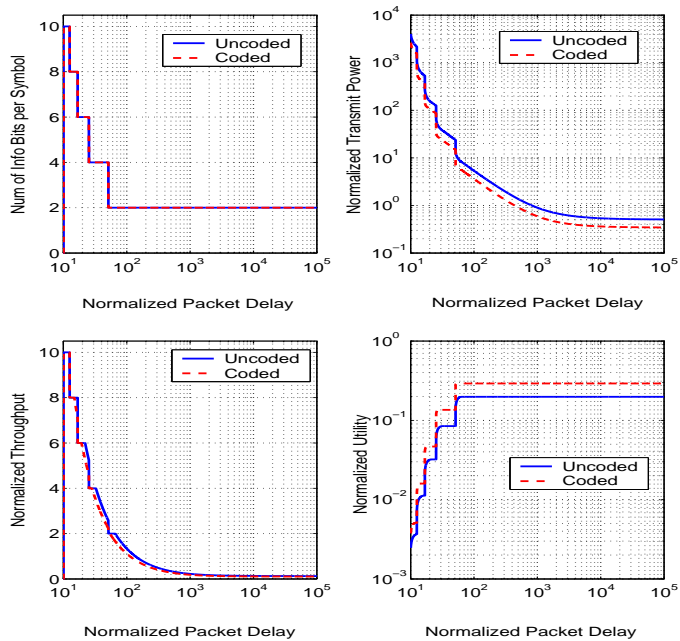}}\vspace{-0.2cm}
\end{center} \caption{Optimum modulation level, transmit power, throughput,
and utility as a function of (normalized) packet delay.}
\label{fig3} \vspace{-0.2cm}
\end{figure}

In this section, we quantify the effect of constellation size on
energy efficiency of a user with a delay QoS constraint. The
source rate (in bps) for the user is assumed to be equal to $0.1B$
where $B$ is the system bandwidth. We further assume that a user
chooses its constellation size, symbol rate, and transmit power
according to its best-response strategy corresponding to the
Pareto-dominant Nash equilibrium (see Section~\ref{pcgd}). For the
coded system, we assume an 8-state convolutional encoder with rate
2/3. The code rate for QPSK is chosen to be 1/2. Fig.~\ref{fig3}
shows the ``optimum" constellation size, transmit power,
throughput, and user's utility as a function of the delay
constraint for both uncoded and coded systems.\footnote{Optimum
here refers to the best-response strategy (i.e., the most
energy-efficient solution).} For all four plots, the packet delay
is normalized by the inverse of the system bandwidth. The
throughput is obtained by multiplying the symbol rate by the
number of (information) bits per symbol, and is normalized by the
system bandwidth. The transmit power and user's utility are also
normalized by $\hat{h}$ and $B\hat{h}$, respectively.
Let us for now focus on the uncoded system. When the delay
constraint is large, QPSK is able to accommodate the delay
requirement and hence is chosen by the user. As the delay
constraint becomes tighter, the user increases its symbol rate and
also raises the transmit power to keep the output SIR at
$\gamma_b^*=9.1$dB. Eventually, a point is reached where the
spectral efficiency of QPSK is not enough to accommodate the delay
constraint. In this case, the user jumps to a higher-order
modulation (i.e., 16-QAM) and the process repeats itself. The
trends are similar for the coded system except that, due to coding
gain, the required transmit power is smaller for the coded system.
This results in an increase in the user's utility (energy
efficiency).

\section{Conclusions}\label{conclusion}

We have studied the effect of modulation order on energy
efficiency of wireless networks using a game-theoretic framework.
Focusing on M-QAM modulation, we have proposed a non-cooperative
game in which each user chooses its strategy in order to maximize
its energy efficiency while satisfying its delay QoS constraint.
The actions open to the users are the choice of the transmit
power, transmit symbol rate and constellation size. The
best-response strategies and the Nash equilibrium solution for the
proposed game have been derived. Using our non-cooperative
game-theoretic framework, the tradeoffs among energy efficiency,
delay, throughput and constellation size have also been studied
and quantified. In addition, we have included the effects of TCM
and have shown that, as expected, coding increases energy
efficiency.


\begin{thebibliography}{10}
{\scriptsize{
\bibitem{WebbSteele}
W.~T. Webb and R.~Steele, ``Variable rate {QAM} for mobile radio,''
{\em IEEE
  Trans. on Commun.}, vol.~43, pp.~2223--2230, July 1995.

\bibitem{GoldsmithChua97}
A.~J. Goldsmith and S.-G. Chua, ``Variable-rate variable-power
{MQAM} for
  fading channels,'' {\em IEEE Trans. on Commun.}, vol.~45,
  pp.~1218--1230, October 1997.

\bibitem{GoldsmithChua98}
A.~J. Goldsmith and S.-G. Chua, ``Adaptive coded modulation for
fading
  channels,'' {\em IEEE Trans. on Commun.}, vol.~46, pp.~595--602,
  May 1998.

\bibitem{Yoo05}
T.~Yoo, R.~J. Lavery, A.~J. Goldsmith, and D.~J. Goodman,
``Throughput
  optimization using adaptive techniques.'' To appear in the \emph{IEEE
  Commun. Letters}.

\bibitem{Cui05}
S.~Cui, A.~J. Goldsmith, and A.~Bahai, ``Energy-constrained
modulation
  optimization,'' {\em IEEE Trans. on Wireless Commun.}, vol.~5,
  pp.~2349--2360, September 2005.


\bibitem{MeshkatiTCOM}
F.~Meshkati, H.~V. Poor, S.~C. Schwartz, and N.~B. Mandayam, ``An
  energy-efficient approach to power control and receiver design in wireless
  data networks,'' {\em IEEE Trans. on Commun.}, vol.~52,
  pp.~1885--1894, November 2005.

\bibitem{MeshkatiISIT}
F.~Meshkati, H.~V. Poor, and S.~C. Schwartz, ``A non-cooperative
power control
  game in delay-constrained multiple-access networks,'' {\em Proceedings of the
  IEEE International Symp. on Info. Theory (ISIT)}, Adelaide,
  Australia, September 2005.

\bibitem{MeshkatiDelayTCOM}
F.~Meshkati, H.~V. Poor, S.~C. Schwartz, and R.~Balan,
``Energy-efficient
  resource allocation in wireless networks with quality-of-service
  constraints.'' preprint, Princeton University, 2005.

\bibitem{GoldsmithBook}
A.~J. Goldsmith, {\em Wireless Communications}.
\newblock Cambridge University Press, New York, NJ, 2005.

\bibitem{MeshkatiModJournal}
F.~Meshkati, A.~J. Goldsmith, H.~V. Poor, and S.~C. Schwartz,
  ``Energy-efficient modulation in CDMA networks with delay QoS constraints.''
  preprint, Princeton University, 2006.

\bibitem{Holliday02}
T.~Holliday, A.~J. Goldsmith, and P.~Glynn, ``Wireless link
adaptation
  policies: {QoS} for deadline constrained traffic with imperfect channel
  estimates,'' {\em Proc. of the IEEE International Conf. on
  Commun. (ICC)}, pp.~3366--3371, New York, NY, April/May 2002.

\bibitem{Ungerboeck}
G.~Ungerboeck, ``Trellis-coded modulation with redundant signal
sets: Parts {I}
  \& {II},'' {\em IEEE Commun. Magazine}, vol.~25, pp.~5--21, February
  1987.
}}
\end{thebibliography}

\end{document}